Winter 1-1-2019

# Mapping of Publications Productivity on Journal of Documentation 1989-2018: A Study Based on Clarivate Analytics – Web of Science Database

Muneer Ahmad
muneerbangroo@gmail.com

Sadik Batcha M

# Mapping of Publications Productivity on Journal of Documentation 1989-2018: A Study Based on Clarivate Analytics – Web of Science Database

**Muneer Ahmad[1] Dr. M Sadik Batcha[2]**
[1]*Research Scholar, Department of Library and Information Science, Annamalai University, Annamalai nagar, muneerbangroo@gmail.com*
[2]*Associate Professor and Head, Library and Information Science Wing, DDE, Annamalai University, Annamalai nagar, msbau@rediffmail.com*

**Abstract**

The Present Study analyzed research productivity in Journal of Documentation (JDoc) for a period of 30 years between 1989 and 2018. Web of Science database a service from Clarivate Analytics has been used to download citation and source data. Bibexcel and Histcite application software have been used to present the datasets. Analysis part focuses on the parameters like citation impact at local and global level, influential authors and their total output, ranking of contributing institutions and countries. In addition to this scientographical mapping of data is presented through graphs using VOSviewer software mapping technique.

**Introduction**

The Present paper analyzed the scientometric examination of research papers contributed in Journal of Documentation (JDoc), which is one of the longest established academic journals in library and information science , providing a distinctive focus on frameworks, models, theories, concepts, and philosophies associated to documents and documented knowledge. This peer-reviewed journal is included in both Clarivate Analytics' SSCI and Scopus and its inception year is 1945. The journal is published by Emerald Group Publishing. The main addressees for the journal are policy-makers, researchers, scholars, and educators in information related areas. It is published quarterly between 1945 and 1996, intensifying to five issues per year between 1997 and 1999. Since 2000, it is available bimonthly. In this study scientometric mapping technique was applied to all the articles published in the JDoc in the period of 30 years.

**Review of Literature**

(S. Batcha, 2018) discussed thoroughly about scientometric output of cardiovascular disease of SAARC countries and offers a powerful set of methods and measures for studying the structure and process of research communication. The paper examines the research trend, authorship, collaborative pattern and activity index of five SAARC Countries regarding the disease which amounts to about 24.8% of deaths in SAARC countries. The result of the paper reveals that India is a leader country among SAARC nations having major research output followed by Pakistan in cardiovascular disease research. The paper also deliberated that USA, England and Australia are the top collaboration countries which has done collaboration with SAARC nations.

(Khan, 2001) attained the scientometric analysis of the DESIDOC Journal of Library and Information Technology from 2010 to 2014. The data from website of journal analysed covers mainly the number of articles, authorship pattern, geographical distribution and types of documents cited. The research reveals that out of 307 contributions, 119 (38.76%) are contributed by single authors while the remaining 188 (61.24%) have joint authors. Furthermore the research finds that maximum contributions are from India.

(M.Sadik Batcha & Muneer Ahmad, 2017b) analysed comparative analysis of Indian Journal of Information Sources and Services (IJISS) and Pakistan Journal of Library and Information Science (PJLIS) during 2011-2017 and studied various aspects like year wise distribution of papers, authorship pattern & author productivity, degree of collaboration pattern of Co-Authorship , average length of papers , average keywords,etc and found 138(94.52%) of contributions from IJISS were made by Indian authors and similarly 94(77.05) of contributions from PJLIS were done by Pakistani authors. Papers by Indian and Pakistani Authors with Foreign Collaboration are minimal (1.37% of articles) and (4.10% of articles) respectively.

(Sab, Kumar, & Biradar, 2018) has studied chemical science research Indian and global output from 2002 and 2016 and applied different data sets for application of different indicators, publications, growth rate, areas of research and discussing its media of communication, strength and weakness in the areas of research, quality of research output, nature of highly cited papers, collaboration national as well as international. It finds chemical science output of India during 2002 -2016 was 5.46% which has increased from 3.94% in 2002 to 6.99% in 2016.

(M. S. Batcha, Jahina, & Ahmad, 2018) has examined scientometric analysis of the DESIDOC Journal and analyzed the pattern of growth of the research output published in the journal, pattern of authorship, author productivity, and, subjects covered to the papers over the period (2013-2017). It found that 227 papers were published during the period of study (2001-2012). The maximum numbers of articles were collaborative in nature. The subject concentration of the journal noted was Scientometrics. The maximum numbers of articles (65 %) have ranged their thought contents between 6 and 10 pages.

(Garg & Tripathi, 2018) study highlights the review of 902 published articles in terms of various disciplines and bibliometric aspects are discussed in these articles. The analysis of data indicates that the share of theoretical studies using mathematical and statistical tools which are missing in the earlier period (1970-1994) has shown an increasing trend during 1995-2014 and the study indicates that the field of medicine as a discipline has received the highest attention as compared to other fields.

(M.Sadik Batcha & Muneer Ahmad, 2017a) conducted scientometric analysis of 146 research articles published in Indian journal of Information Sources and Services (IJISS). The number of contributions, authorship pattern & author productivity, average citations, average length of articles, average keywords and collaborative papers was analyzed. Out of 146 contributions, only 39 were single authored and rest by multi authored with degree of collaboration 0.73 and week collaboration among the authors. The study concluded that the author productivity was 0.53 and was dominated by the Indian authors.

(Dutt & Nikam, 2016) examined 10905 global publication output in solar cell research for five years from 1991 to 2010 indexed in web of science. The major contributors are countries like USA and China and Indian is positioned at 6$^{th}$ place. The majority of output coming out form academic institutions and major concentration has been on aspects of research pertaining to chemical science. Chinese Academy of Sciences outperformed other academic institutions; however its impact was relatively lower than other prolific institutions.

(Biljecki, 2016) examined a set of 12436 papers published in 20 GIScience journals in the gap of 2000-2014 and studied patterns and trends and its comprehensive scientometric study focuses on multiple aspects like output volume, citations, national output and efficiency, collaboration, altmetrics, authorship, and length of articles. The notable observation are that 5% countries contribute 76% global GIScience output, a paper published 15 years before received a median of 12 citations and the share of global collaborations in GIScience has more than tripled from the year 2000 onwards(31% papers has multiple authors from multiple countries in 2014 and it increased from 10% in 2014) .

(Ahmad, Batcha, Wani, Khan, & Jahina, 2017) explored scientometric analysis of the Webology Journal. The paper analyses the pattern of growth of the research output published in the journal, pattern of authorship, author productivity, and subjects covered to the papers over the period (2013-2017). It was found that 62 papers were published during the period of study (2013-2017). The maximum numbers of articles were collaborative in nature. The subject concentration of the journal noted was Social Networking/Web 2.0/Library 2.0 and Scientometrics or Bibliometrics. Iranian researchers contributed the maximum number of articles (37.10%). The study applied standard formula and statistical tools to bring out the factual results.

(M. S. Batcha, 2017) analysed the research publication output in the field of robotic technology and shows that the robotic technology is a progressive field increasing the publication output from single digit to 513 year after year during the period from 1990 to 2016. The results shows that developing countries like USA, UK and Germany gives the most output on robotic technology related research. Yet major proportion of contribution (36.30%) is from USA. English language is the most preferred for the research amounting (87.70%) followed by German. The Prolific authors in the field of robotic technology are highly found from USA among them the contribution by Bloss R is appreciable and author from Japan, Dario P competes with more number of publications in the study.

**Objectives**

The main objective of the study is to consider on the mapping of 2220 articles published by the Journal of Documentation (JDoc) during the period of 1989 – 2018 and the specific objectives are to identify and carry out the following factors

- To examine the annual publications output of Journal of Documentation.
- To gauge publication density through mapping of top 30 authors, countries and institutions based on their number of research papers.
- Find out the top 30 prolific authors, institutions and countries.

## Data Source and Methodology

The data for the present study were downloaded from the Clarivate analytics-Web of Science database in December 2018. A total of 2220 research publications was downloaded from 1989-2018. The data downloaded were enhanced with different parameters like title, authors, years, countries, and research institutions. Furthermore, the downloaded data were analyzed by using Bibexcel, Histcite, and Vosviewer software applications.

Table – I: Details of the Important Points of the Data Sample During 1989 to 2018

| S.No. | Details about Sample | Observed Values |
|---|---|---|
| 1 | Duration | 1989-2018 |
| 2 | Collection Span | 30 Years |
| 3 | Total No. of Records | 2220 |
| 4 | Total No. of Authors | 1626 |
| 5 | Frequently Used Words | 3888 |
| 6 | Document Types | 12 |
| 7 | Languages | 2 |
| 8 | Contributing Countries | 64 |
| 9 | Contributing Institutions | 633 |
| 10 | Institutions with Sub Division | 1112 |
| 11 | Total Cited References | 38798 |
| 12 | Total Local Citation Scores | 1945 |
| 13 | Total Global Citation Scores | 18037 |
| 14 | H-Index | 56 |

## Discussion and Result

### Evaluate the Annual Output of Publications

The table II reveals that the numbers of research documents published from 1989 to 2018 are gradually increased. According to the publication output from the table II the year wise distribution of research documents, 2001 has the highest number of research documents 109 (4.91%) with 82 (4.22%) of total local citation score and 882 (4.89%) of total global citation score values and being prominent among the 30 years output and it stood in first rank position. The year 1997 has 97 (4.37%) research documents and it stood in second position with 87 (4.47%) of total local citation score and 1067 (5.92%) of total global citation score were scaled. It is followed by the year 2002 with 93 (4.19 %) of records and it stood in third rank position along with 72(3.70%) of total local citation score and 755 (4.19%) of total global citation score measured. The year 1999 has 87 (3.92%) research documents and it stood in fourth position with 117 (6.02%) of total local citation score and 1159(6.43%) of total global citation score were scaled. It is noticed that the increase in publications may not create impact on citation score yet the quality matters on total local citation scores and on total global citation scores. Graph number one mull over the year wise publications and depicts the citation score. It clearly indicates on the fact that the increased publication rate is not bringing the increased citation rate.

Table – II: Annual Distribution of Publications and Citations

| S.No. | Publication Year | Records | % | Rank | TLCS | % | Rank | TGCS | % | Rank |
|---|---|---|---|---|---|---|---|---|---|---|
| 1 | 1989 | 71 | 3.20 | 17 | 59 | 3.03 | 18 | 569 | 3.15 | 18 |
| 2 | 1990 | 71 | 3.20 | 17 | 43 | 2.21 | 22 | 375 | 2.08 | 22 |
| 3 | 1991 | 80 | 3.60 | 8 | 38 | 1.95 | 25 | 312 | 1.73 | 24 |
| 4 | 1992 | 79 | 3.56 | 9 | 83 | 4.27 | 8 | 491 | 2.72 | 19 |
| 5 | 1993 | 74 | 3.33 | 11 | 76 | 3.91 | 10 | 785 | 4.35 | 9 |
| 6 | 1994 | 72 | 3.24 | 15 | 17 | 0.87 | 29 | 151 | 0.84 | 27 |
| 7 | 1995 | 75 | 3.38 | 10 | 39 | 2.01 | 24 | 386 | 2.14 | 21 |
| 8 | 1996 | 85 | 3.83 | 5 | 61 | 3.14 | 17 | 570 | 3.16 | 17 |
| 9 | 1997 | 97 | 4.37 | 2 | 87 | 4.47 | 7 | 1067 | 5.92 | 3 |
| 10 | 1998 | 84 | 3.78 | 6 | 72 | 3.70 | 12 | 888 | 4.92 | 6 |
| 11 | 1999 | 87 | 3.92 | 4 | 117 | 6.02 | 2 | 1159 | 6.43 | 1 |
| 12 | 2000 | 73 | 3.29 | 12 | 58 | 2.98 | 19 | 699 | 3.88 | 13 |
| 13 | 2001 | 109 | 4.91 | 1 | 82 | 4.22 | 9 | 882 | 4.89 | 7 |
| 14 | 2002 | 93 | 4.19 | 3 | 72 | 3.70 | 12 | 755 | 4.19 | 12 |
| 15 | 2003 | 73 | 3.29 | 12 | 107 | 5.50 | 3 | 999 | 5.54 | 5 |
| 16 | 2004 | 61 | 2.75 | 26 | 49 | 2.52 | 21 | 862 | 4.78 | 8 |
| 17 | 2005 | 84 | 3.78 | 6 | 137 | 7.04 | 1 | 1024 | 5.68 | 4 |
| 18 | 2006 | 70 | 3.15 | 21 | 70 | 3.60 | 14 | 764 | 4.24 | 10 |
| 19 | 2007 | 67 | 3.02 | 23 | 103 | 5.30 | 4 | 763 | 4.23 | 11 |
| 20 | 2008 | 71 | 3.20 | 17 | 62 | 3.19 | 16 | 1159 | 6.43 | 1 |
| 21 | 2009 | 72 | 3.24 | 15 | 64 | 3.29 | 15 | 624 | 3.46 | 15 |
| 22 | 2010 | 59 | 2.66 | 28 | 100 | 5.14 | 5 | 643 | 3.56 | 14 |
| 23 | 2011 | 67 | 3.02 | 23 | 91 | 4.68 | 6 | 593 | 3.29 | 16 |
| 24 | 2012 | 55 | 2.48 | 29 | 73 | 3.75 | 11 | 426 | 2.36 | 20 |
| 25 | 2013 | 55 | 2.48 | 29 | 31 | 1.59 | 27 | 360 | 2.00 | 23 |
| 26 | 2014 | 62 | 2.79 | 25 | 56 | 2.88 | 20 | 265 | 1.47 | 25 |
| 27 | 2015 | 70 | 3.15 | 21 | 38 | 1.95 | 25 | 221 | 1.23 | 26 |
| 28 | 2016 | 60 | 2.70 | 27 | 40 | 2.06 | 23 | 150 | 0.83 | 28 |
| 29 | 2017 | 73 | 3.29 | 12 | 18 | 0.93 | 28 | 83 | 0.46 | 29 |
| 30 | 2018 | 71 | 3.20 | 17 | 2 | 0.10 | 30 | 12 | 0.07 | 30 |
| | **Total** | **2220** | **100.00** | | **1945** | **100.00** | | **18037** | | |

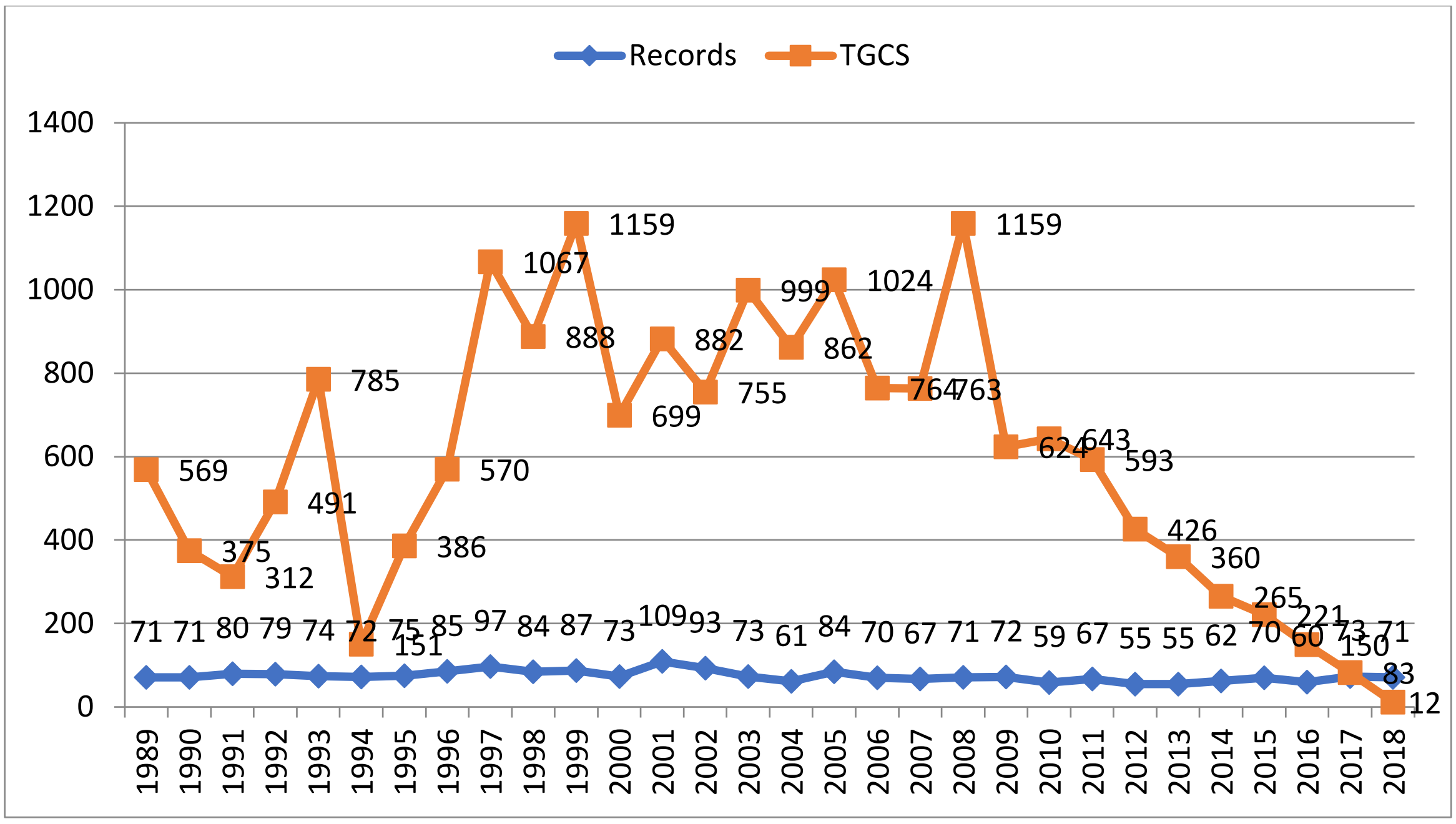

Graph 1: Annual Distribution of Publications and Citations

**Analysis of the Publication Output of Top 30 Authors**

Table III and figure 1 displays the ranking of authors of research articles. In the rank analysis the authors who have published more than 10 articles or more are considered into account to avoid a long list. It was observed that there is total of 1626 authors for 2220 records and it shows the top 30 most productive authors during 1989-2018. Bawden D published 99 (4.46%) articles with 358 TGCS articles, followed by Line MB 55 (2.48%) with 17 TGCS articles, Oppenheim C 37 (1.67%) with 419 TGCS articles, Cronin B 33 (1.49%) with 395 TGCS articles, Vickery B 27 (1.22%) with 88 TGCS article, Davenport E 23 (1.04%) with 77 TGCS articles, other authors have contributed less than 1% during the period of study. The data set clearly depicts that no matter how many publications that an author brings out yet the quality publications alone shows impact in the form of total local citations score and total global citations score. It is found that the ranked contributors are from City University London, University Sheffield Department of information studies, Royal School Library and Information Science and Queens University Belfast.It could be identified that the authors' wise analysis the following authors Bawden D, Line MB, Oppenheim C, Cronin B, Vickery B, and Davenport E, were identified the most productive authors based on the number of research papers published. The data set puts forth that the authors Ellis D with 937 citations, Hjorland B with 673 citations, Ford N with 442 citations and Oppenheim C with 419 citations.

Table III: Publication output of Top 30 Authors and Citation Score

| Rank | Authors | Records | % | TGCS |
|---|---|---|---|---|
| 1 | Bawden D | 99 | 4.46 | 358 |
| 2 | Line MB | 55 | 2.48 | 17 |
| 3 | Oppenheim C | 37 | 1.67 | 419 |
| 4 | Cronin B | 33 | 1.49 | 395 |
| 5 | Vickery B | 27 | 1.22 | 88 |
| 6 | Davenport E | 23 | 1.04 | 77 |
| 7 | Savolainen R | 20 | 0.90 | 317 |
| 8 | Hjorland B | 19 | 0.86 | 673 |
| 9 | Pors NO | 19 | 0.86 | 6 |
| 10 | Willett P | 19 | 0.86 | 192 |
| 11 | Ford N | 18 | 0.81 | 442 |
| 12 | Vakkari P | 18 | 0.81 | 383 |
| 13 | Warner J | 17 | 0.77 | 97 |
| 14 | Meadows J | 16 | 0.72 | 7 |
| 15 | Urquhart C | 16 | 0.72 | 68 |
| 16 | Cawkell T | 15 | 0.68 | 2 |
| 17 | Hannabuss S | 15 | 0.68 | 0 |
| 18 | Robinson L | 15 | 0.68 | 117 |
| 19 | Ellis D | 14 | 0.63 | 937 |
| 20 | Marcella R | 14 | 0.63 | 79 |
| 21 | Nicholas D | 14 | 0.63 | 247 |
| 22 | Rousseau R | 14 | 0.63 | 167 |
| 23 | Rowley J | 14 | 0.63 | 75 |
| 24 | Thelwall M | 14 | 0.63 | 301 |
| 25 | Ashford J | 13 | 0.59 | 4 |
| 26 | Corrall S | 11 | 0.50 | 0 |
| 27 | Huntington P | 11 | 0.50 | 234 |
| 28 | Sturges P | 11 | 0.50 | 25 |
| 29 | Zumer M | 11 | 0.50 | 88 |
| 30 | Bade D | 10 | 0.45 | 1 |

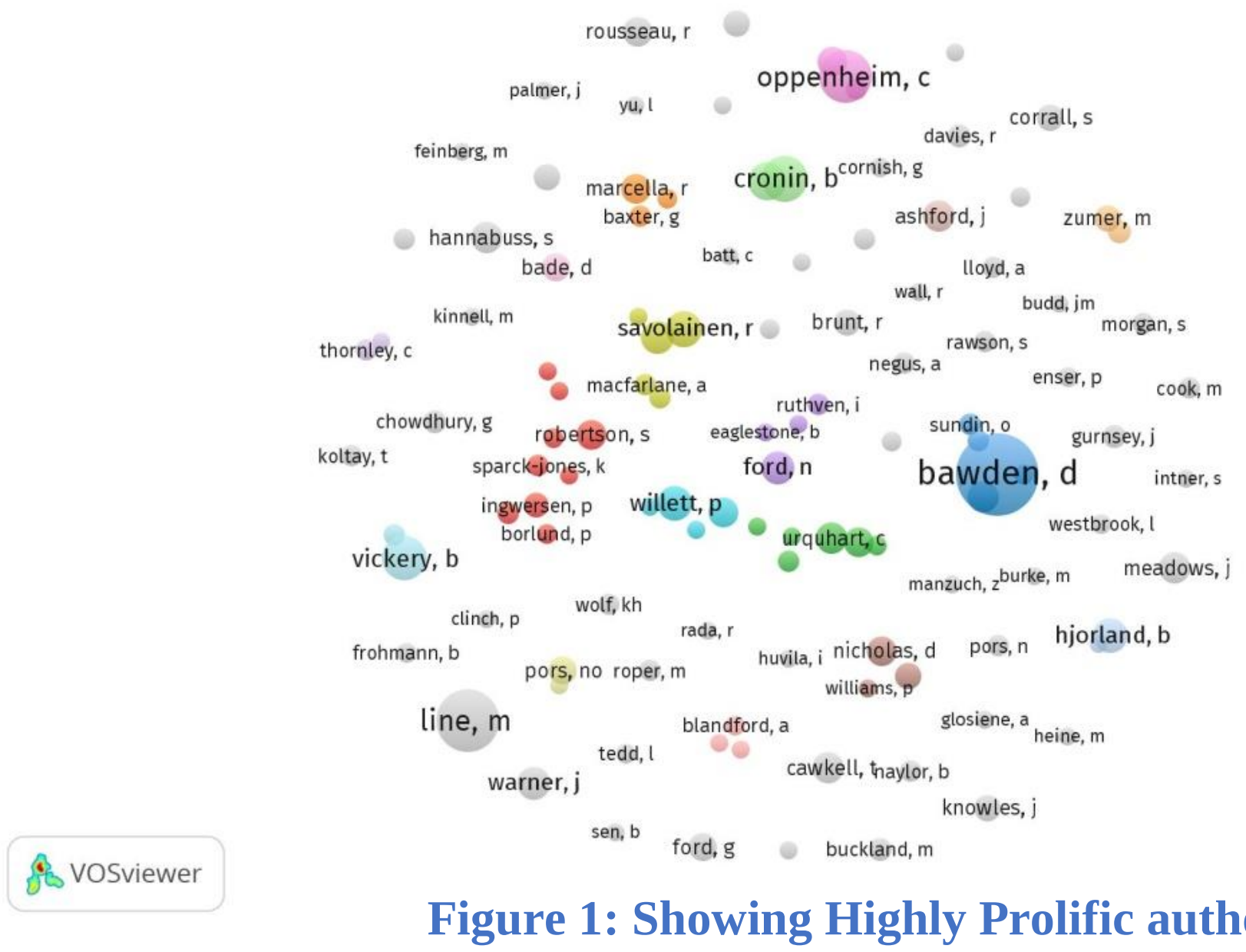

Figure 1: Showing Highly Prolific authors

**ANALYSIS OF THE PUBLICATION OUTPUT OF TOP 30 INSTITUTIONS**

The individualities of 30 most productive institutions were analyzed in this part, Institutions which published more than 14 and above publications have considered as highly productive institutions. Table IV summarizes articles, the global citation score, local citation score and average author per paper of the publications of these institutions.

In total, 633 institutions, including 1112 subdivisions published 2220 research papers during 1989 – 2018. The topmost thirty prolific institutions involved in this research have published 14 and more research articles. The mean average is 3.51 research articles per institution. Out of 633 institutions, top 30 institutions published 1275 (57.43%) research papers and the rest of the institution published 945 (42.57%) research papers respectively. Based on the number of published research records the institutions are ranked.

The institution “City University London” holds the first rank and the institution published 111 (5%) research papers with 131 local and 1352 global citation scores, the average citation per paper is 12.18. The second rank holds by “University Sheffield” the institution published 97 (4.37%) research papers with 301 local and 2643 global citation scores, the average citation per paper is 27.25. The “Queens University Belfast” holds the 3rd rank, the institution published 59 (2.66%) research papers with 13 local and 101 global citation scores, the average citation per paper is 1.71. The “University Loughborough” holds the 4th rank, the institution published 59 (2.66%) research papers with 33 local and 424 global citation scores, the average citation per paper is 7.19. The “Royal School Library & Information Science” holds the 5th rank; the institution published 56 (2.52%) research papers with 148 local and 1337 global citation scores, the average citation per paper is 23.88. It is clear from the analysis that following institutions City University London, University Sheffield, Queens University Belfast, University Loughborough and Royal School Library and Information Science were identified the most productive institutions based on the number of research papers published and besides these institutions some unidentified institutions categorised under “Unknown” institutions amounted 324 records having 33 and 118 local and global citation scores respectively. University Sheffield (27.25), Royal School Library & Information Science (23.88), University Western Ontario (22.65), University Toronto (19.13) and University Tampere (14.27) are the institutions with high ACPP indicating the quality work with high citation impact hence they can be recognized as the most industrious institutions based on the annual citation per paper received in terms of publications.

Table IV: Ranking of Institutions and their Research Performance

| S.No. | Institution | Records | % | TLCS | TGCS | ACPP |
|---|---|---|---|---|---|---|
| 1 | City University London | 111 | 5.00 | 131 | 1352 | 12.18 |
| 2 | University Sheffield | 97 | 4.37 | 301 | 2643 | 27.25 |
| 3 | Queens University Belfast | 59 | 2.66 | 13 | 101 | 1.71 |
| 4 | University Loughborough | 59 | 2.66 | 33 | 424 | 7.19 |
| 5 | Royal School Library & Information Science | 56 | 2.52 | 148 | 1337 | 23.88 |
| 6 | University Tampere | 51 | 2.30 | 102 | 728 | 14.27 |
| 7 | Indiana University | 49 | 2.21 | 47 | 468 | 9.55 |
| 8 | Loughborough University Technology | 44 | 1.98 | 9 | 142 | 3.23 |
| 9 | University Strathclyde | 44 | 1.98 | 39 | 375 | 8.52 |
| 10 | Robert Gordon University | 39 | 1.76 | 17 | 130 | 3.33 |
| 11 | UCL | 35 | 1.58 | 36 | 456 | 13.03 |
| 12 | University Ljubljana | 26 | 1.17 | 10 | 101 | 3.88 |
| 13 | University Wales | 23 | 1.04 | 4 | 63 | 2.74 |
| 14 | University Boras | 21 | 0.95 | 35 | 161 | 7.67 |
| 15 | University Western Ontario | 20 | 0.90 | 88 | 453 | 22.65 |
| 16 | British Library | 19 | 0.86 | 2 | 3 | 0.16 |
| 17 | Northumbria University | 18 | 0.81 | 5 | 52 | 2.89 |
| 18 | Nanyang Technology University | 17 | 0.77 | 15 | 116 | 6.82 |
| 19 | Aberystwyth University | 16 | 0.72 | 17 | 69 | 4.31 |
| 20 | Lund University | 16 | 0.72 | 30 | 113 | 7.06 |
| 21 | University Bath | 15 | 0.68 | 15 | 79 | 5.27 |
| 22 | University Texas Austin | 15 | 0.68 | 14 | 66 | 4.40 |
| 23 | University Toronto | 15 | 0.68 | 41 | 287 | 19.13 |
| 24 | University Wisconsin | 15 | 0.68 | 14 | 97 | 6.47 |
| 25 | Vilnius State University | 15 | 0.68 | 0 | 6 | 0.40 |
| 26 | Drexel University | 14 | 0.63 | 12 | 42 | 3.00 |
| 27 | Leeds Metropolitan University | 14 | 0.63 | 8 | 43 | 3.07 |
| 28 | Manchester Metropolitan University | 14 | 0.63 | 6 | 83 | 5.93 |
| 29 | Napier University | 14 | 0.63 | 1 | 15 | 1.07 |
| 30 | Unknown | 324 | 14.59 | 33 | 118 | 0.36 |

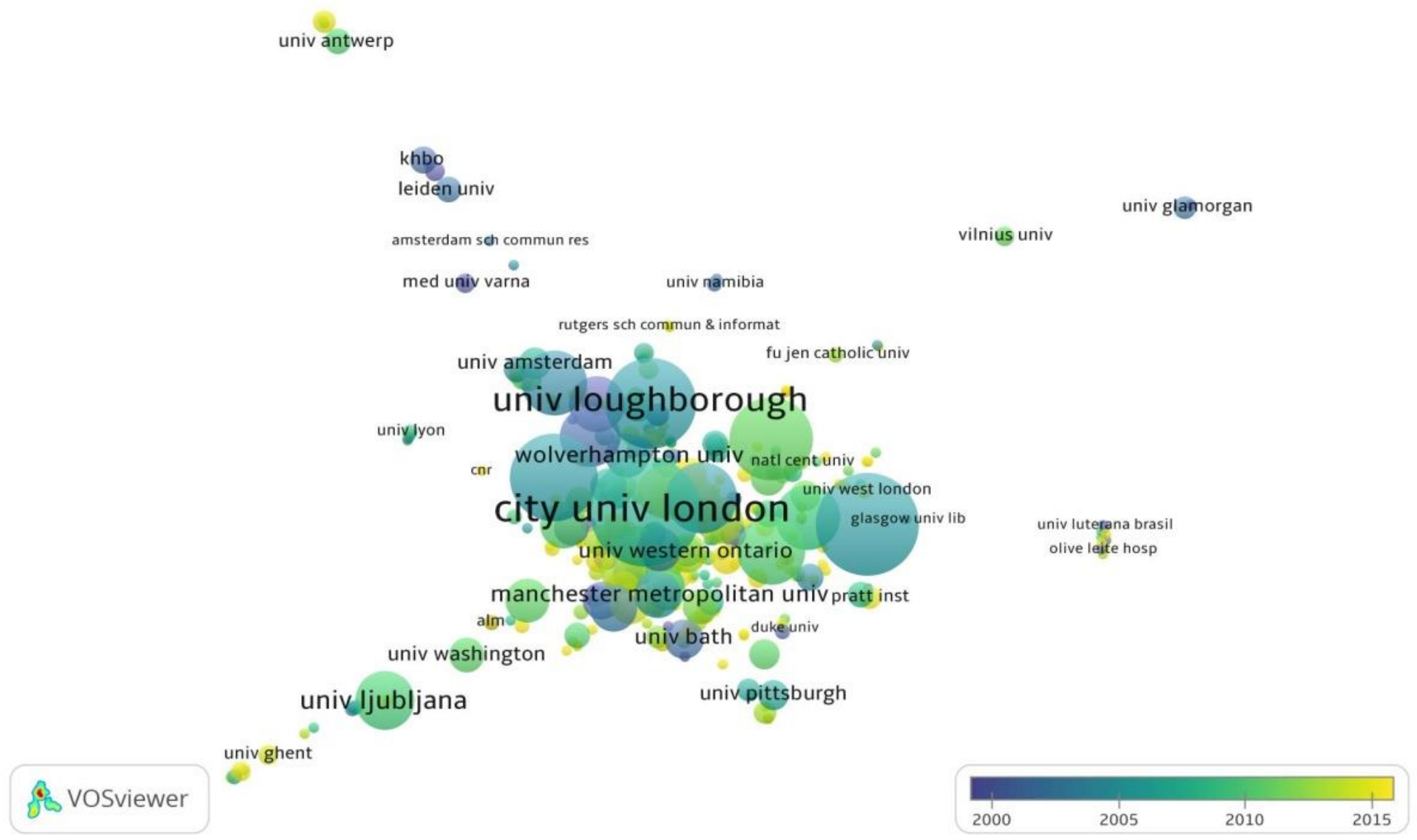

**Figure 2: Collaboration of Institutions and their clusters**

**Analysis of the Publication Output of Top 30 Countries**

Table V and figure 3 displays the publication output of the top thirty countries by number of papers and UK acquired 1st rank among the top thirty countries under consideration with its total global citation score 6864 (40.95%).In all 64 countries participated in research during 1989 and 2018. The countries that rank between 2nd and 30th position are USA, Finland, Canada, Denmark, Sweden, Australia, Peoples Republic of China, Slovenia, Belgium, Spain, Netherlands, Singapore, Norway, Germany, Ireland, Lithuania, Italy, Hungary, Taiwan, Japan, New Zealand, Croatia, France, Iran, South Korea, Switzerland, Brazil. We have found by using this country mapping analysis that there are nodes with clarity of linking between each node, which indicates that there are countries linking and associated with other associated countries. It could be identified that the country wise analysis the following countries UK, USA, Finland, Canada, Denmark, Sweden were identified the most productive country based on the number of research papers published.

Table V: Distribution of the Publication Output of Top 30 Countries

| S.No. | Country | Records | % | TLCS | TGCS |
|---|---|---|---|---|---|
| 1 | UK | 909 | 40.95 | 716 | 6864 |
| 2 | Unknown | 519 | 23.38 | 232 | 1998 |
| 3 | USA | 281 | 12.66 | 289 | 2654 |
| 4 | Finland | 71 | 3.20 | 143 | 1083 |
| 5 | Canada | 58 | 2.61 | 138 | 873 |
| 6 | Denmark | 57 | 2.57 | 133 | 1521 |
| 7 | Sweden | 53 | 2.39 | 90 | 412 |
| 8 | Australia | 47 | 2.12 | 84 | 520 |
| 9 | Peoples R China | 32 | 1.44 | 11 | 199 |
| 10 | Slovenia | 24 | 1.08 | 10 | 80 |
| 11 | Belgium | 22 | 0.99 | 14 | 188 |
| 12 | Spain | 22 | 0.99 | 9 | 161 |
| 13 | Netherlands | 20 | 0.90 | 23 | 301 |
| 14 | Singapore | 18 | 0.81 | 16 | 125 |
| 15 | Norway | 17 | 0.77 | 16 | 154 |
| 16 | Germany | 16 | 0.72 | 12 | 143 |
| 17 | Ireland | 16 | 0.72 | 5 | 28 |
| 18 | Lithuania | 16 | 0.72 | 0 | 7 |
| 19 | Italy | 11 | 0.50 | 7 | 92 |
| 20 | Hungary | 9 | 0.41 | 3 | 37 |
| 21 | Taiwan | 9 | 0.41 | 6 | 76 |
| 22 | Japan | 8 | 0.36 | 3 | 63 |
| 23 | New Zealand | 8 | 0.36 | 9 | 145 |
| 24 | Croatia | 7 | 0.32 | 7 | 51 |
| 25 | France | 7 | 0.32 | 4 | 36 |
| 26 | Iran | 6 | 0.27 | 9 | 92 |
| 27 | South Korea | 5 | 0.23 | 5 | 170 |
| 28 | Switzerland | 5 | 0.23 | 7 | 530 |
| 29 | Brazil | 4 | 0.18 | 1 | 15 |
| 30 | Czech Republic | 4 | 0.18 | 0 | 1 |

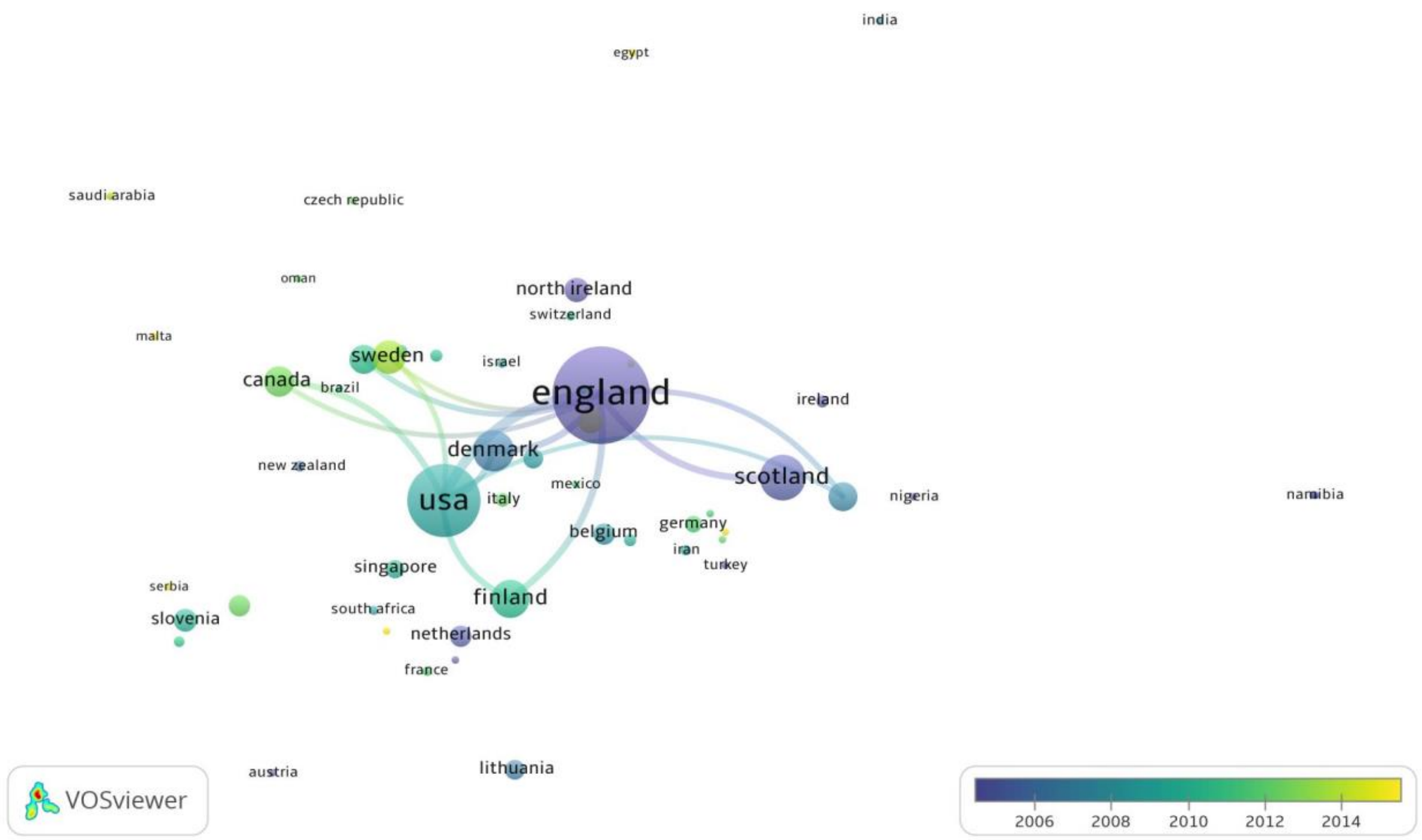

**Figure 3: Showing Ranking of Country wise Distribution**

## Conclusion

The number of papers published in Journal of Documentation (JDoc) has gradually increased during 1989–2018 and the study has shown that 2220 research documents have been published in Journal of Documentation during the period. It could be identified that the author's wise analysis the following authors Bawden D, Line MB, Oppenheim C, Cronin B, Vickery B, and Davenport E, were acknowledged the most prolific authors based on the number of research papers contributed. It could be identified that the institutions wise analysis the following institutions City University London, University Sheffield, Queens University Belfast, University Loughborough and Royal School Library and Information Science were acknowledged the most prolific institutions based on the number of research papers output they published. It could be identified that the country wise analysis the following countries UK, USA, Finland, Canada, Denmark, and Sweden were identified the most productive country based on the number of research papers published.

## References

Ahmad, M., Batcha, M. S., Wani, B. A., Khan, M. I., & Jahina, S. R. (2017). Research Output of Webology Journal ( 2013-2017 ): A Scientometric Analysis. *International Journal of Movement Education and Social Science*, *7*(3), 46–58.

Batcha, M. S. (2017). Research Output Analysis on Robotic Technology : A Scientometric Study. *Indian Journal of Information Sources and Services*, *7*(1), 25–31.

Batcha, M. S., Jahina, S. R., & Ahmad, M. (2018). Publication Trend in DESIDOC Journal of Library and Information Technology during 2013-2017 : A Scientometric Approach. *International Journal of Research in Engineering, IT and Social Sciences*, *08*(04), 76–82.

Batcha, S. (2018). Cardiovascular Disease Research in SAARC Countries: A Scientometric Study. *Informatics Studies*, *5*(4).

Biljecki, F. (2016). A scientometric analysis of selected GIScience journals. *International Journal of Geographical Information Science*, *30*(7), 1302–1335. https://doi.org/10.1080/13658816.2015.1130831

Dutt, B., & Nikam, K. (2016). Scientometric analysis of global solar cell research. *Annals of Library and Information Studies*, *63*(1), 31–41. https://doi.org/10.1152/physiol.00052.2014

Garg, K. C., & Tripathi, H. K. (2018). Bibliometrics and scientometrics in India: An overview of studies during 1995-2014 Part II: Contents of the articles in terms of disciplines and their bibliometric aspects. *Annals of Library and Information Studies*, *65*(March), 7–42. Retrieved from http://nopr.niscair.res.in/bitstream/123456789/44221/3/ALIS 65%281%29 7-42.pdf

Khan, I. (2001). Library Hi Tech News Repec. *Library Hi Tech News*, *33*(7), 1–19.

M.Sadik Batcha & Muneer Ahmad. (2017a). Publication Pattern Analysis in Indian Journal of Information Sources and Services (IJISS): A Scientometric Approach. In *Rejuvenating Libraries for Information Access in the Digital Era* (pp. 1078–1090).

M.Sadik Batcha & Muneer Ahmad. (2017b). Publication Trend in an Indian Journal and a Pakistan Journal : A Comparative Analysis using Scientometric Approach. *Journal of Advances in Library and Information Science*, *6*(4), 442–449. Retrieved from http://jalis.in/pdf/6-4/Sadik.pdf

Sab, M. C., Kumar, P. D., & Biradar, B. S. (2018). Growth of Literature and Measures of Scientific Productivity of Indian Chemical Science Research during 2002-2016. *Journal of Advanced Chemical Sciences*, *04*(01), 525–530. https://doi.org/10.30799/jacs.178.18040101